

\magnification=1200
\baselineskip=22pt plus 1pt minus 1pt
\tolerance=1000
\parskip=0pt
\parindent=25pt

\font\rmbig=cmr12 scaled\magstep1

\font\bxBig=cmbx12  scaled\magstep2

\font\medsl=cmsl8

\def\normal{\baselineskip=22pt plus 1pt minus 1pt}

\def\ep{\epsilon}
\def\eps{\varepsilon^{\mu\nu\rho}}
\def\d{\partial}
\def\la{\raise.16ex\hbox{$\langle$} }
\def\ra{\raise.16ex\hbox{$\rangle$} }
\def\go{\rightarrow}

\def\display{ \displaystyle \strut }

\def\psibar{ \psi \kern-.65em\raise.6em\hbox{$-$} }
\def\Dbar{ D \kern-.8em\raise.65em\hbox{$-$} }

\def\ha{ \hat a }
\def\ta{ \tilde a }

\def\N{ \kappa }

\def\x{ {\bf x} }
\def\y{ {\bf y} }

\def\r{ {\bf r} }

\nopagenumbers

\baselineskip=14pt

\centerline{\rmbig Theoretical Physics Institute}
\centerline{\rmbig University of Minnesota}
\vskip .8 cm

\baselineskip=10pt
\line{September 1992 \hfil
  \vtop{ \hsize=4cm  TPI-MINN-92/41-T

                      UMN-TH-1105/92} }

\vskip 1.8cm

\baselineskip=22pt
\centerline{\bxBig Operator Algebra}
\centerline{\bxBig in Chern-Simons Theory on a Torus}

\vskip 1.8cm

\baselineskip=13pt

\centerline{\bf Choon-Lin Ho}
\centerline{\medsl Theoretical Physics Institute, University of Minnesota,
  Minneapolis, MN 55455, U.S.A.,  and}
\centerline{\medsl Department of Physics, Tamkang University, Tamsui,
Taiwan 25137, R.O.C.\footnote{$^\dagger$}{Permanent address}}

\vskip .2cm
\centerline{and}
\vskip .2cm

\centerline{\bf Yutaka Hosotani}
\centerline{\medsl School of Physics and Astronomy, University of Minnesota,
  Minneapolis, MN 55455, U.S.A.}

\vskip 1.5cm

\baselineskip=16pt

\centerline{\bf Abstract}

\midinsert \narrower
We consider Chern-Simons gauge theory on a torus with both nonrelativistic and
relativistic matter.  It is shown that the Hamiltonian and  two total
momenta commute among themselves  only in the physical
Hilbert space.  We also discuss  relations among degenerate physical states,
degenerate vacua, and the existence of multicomponent Schr\"odinger
wavefunctions. \endinsert

\vfill\eject

\footline={\hss\tenrm\folio\hss}


\pageno=2

\normal
\parindent=25pt

Chern-Simons field theory with matter coupling have attracted intense
interest in recent years, owing to its relevance to condensed matter systems
such as  quantum Hall systems, and possibly  high $T_c$
superconductors [1].  Setting aside such physical
applicability, Chern-Simons theory is in itself very interesting in view of
its rich and beautiful mathematical structures.  As such, various
aspects of this theory deserve a careful study.  Much of recent efforts have
been directed towards the issue of
consistent quantisation of the theory [2,3].

While the majority of works in the field is concerned with planar systems,
Chern-Simons field theory on compact Riemann surfaces has also captured
considerable interests [4--14].  It has even richer structures, which, being
topological in nature, are absent in planar system.  Among them are the
multicomponent structure of  many-body wavefunctions [9,11,14] and the
degeneracy of physical states [5].   Moreover, the analysis on a torus is
mathematically rigorous, being free from infrared divergences and ambiguity in
boundary conditions at space infinity on a plane.

We extend our previous analysis [14], examining algebraic relations among
various operators, especially the Hamiltonian $H$ and  total momenta $P^k$,
with an eye on whether translation invariance is maintained or broken
in the presence of matter.  We shall
also discuss a possible link between degeneracy of physical states and the
multicomponent structure of wavefunctions.

It has been argued by Chen et al.[15] that the microscopic translation
invariance  of the anyon superconductivity model is broken in the mean field
approximation, and is restored in the random phase approximation thanks to the
presence of the phonon mode.  Our consideration is at the microscopic level.
We shall show that $H$ and $P^k$ do not commute with each other as operators,
whereas they do commute in the physical Hilbert space.  Therefore, if an
effective theory is formulated in terms of physical excitations,
the translation invariance must be maintained manifestly in each mode.
Similar arguments have been given by Iengo, Lechner, and Li [13] on a torus,
and
by Banerjee [16] on a plane.  Our argument, however, differs from theirs in
detail.

We consider two models, Chern-Simons gauge theories on a torus with a
non-relativistic matter field and with a relativistic Dirac field.  We shall
find algebraic relations  universal in both theories.

We first analyse the nonrelativistic case.  The Lagrangian is given by
$${\cal L}= {\N \over 4\pi} ~\eps a_\mu \d_\nu a_\rho
   + {\cal L}_{\rm matter} \eqno(1) $$
where
$$
{\cal L}_{\rm matter}=
{i\over 2} \Big\{ \psi^\dagger  D_0 \psi - (D_0 \psi )^\dagger \psi \Big\}
 - {1\over 2m} (D_k\psi)^\dagger (D_k\psi) ~~,   \eqno(2)  $$
$D_0 = \d_0 +  i a_0$, and $D_k = \d_k - i a^k$.
 Consistency of the theory requires that the coefficient of the
Chern-Simons term $\N$ be a fractional ratio, $\N=N/M$, where  $N$ and $M$ are
coprime integers [7,8].
The funamental domain of the torus is given by $0\leq x_j \leq L_j, j=1,2$.
 Boundary conditions of the fields are then [4,14]
$$\eqalign{
a_\mu[T_j(x)] &= a_\mu[x] + \d_\mu \beta_j (x) ~,\cr
\psi[T_j(x)] &= e^{-i\beta_j(x) } ~ \psi(x)~, \cr}
    \eqno(3) $$
where $T_j : ~ x_j \rightarrow x_j + L_j$ ($j=1,2$).
That is,  the fields return to their original values up to gauge
transformations after translations along non-contractible loops.
The requirement of smoothness of the field
operator $\psi (x)$ in the covering space, $\psi[T_1\cdot T_2 (x)] =
\psi[T_2\cdot T_1 (x)] $, leads to quantisation of the Chern-Simons flux,
$\Phi=\int d{\bf x} ~f_{12} = 2\pi m$ ($m$ : integers), and a constraint on the
$\beta_j$'s:
$\big\{ \beta_1 (T_2x) - \beta_1 (x) \big\} -
\big\{ \beta_2(T_1x) -\beta_2 (x) \big\} = -2\pi m$.
Typical $\beta_j$'s which solve this constraint  are $\beta_j
(x)=-\epsilon^{jk}
\pi m x_k / L_k$, which will be taken in the rest of the paper.

Canonical energy-momentum tensors, $T^{\mu\nu}_c$, derived from (1) and (2)
are not gauge-invariant, and therefore are not well defined on a torus in view
of  the boundary conditions (3).  The gauge-invariant  energy-momentum tensors
$T^{\mu\nu}_I$ are obtained by adding the term $(\N / 4\pi)
{}~\d_\rho \left(\ep^{\rho\mu\sigma} a_\sigma a^\nu\right)$ to
$T^{\mu\nu}_c$ and making use of the equations of motion [3]:
$$\eqalign{
&T^{\mu\nu}_I
= \left[
 {\display {i\over \,2\,} \{ \psi^\dagger D^\nu\psi
      - (D^\nu\psi)^\dagger \psi \} \atop
  \display -{1\over 2m} \big\{ (D^k\psi)^\dagger D^\nu\psi
 + (D^\nu\psi)^\dagger D^k\psi \big\} } \right] \cr
\noalign{\kern 5pt}
&\hskip .2cm - g^{\mu\nu} \bigg\{ {i\over \,2\,} \Big( \psi^\dagger D^0\psi
              - (D^0\psi)^\dagger \psi \Big)
  -{1\over 2m} (D^k\psi )^\dagger D^k\psi \bigg\} ~~.\cr}  \eqno(4)$$
where the upper (lower) entries in the square bracket give the $\mu=0$
($\mu=k$)
component of $T^{\mu\nu}_I$.
{}From (4) we obtain the Hamiltonian and  total momentum operators
$$\eqalign{
H &= {1\over 2m}~\int d{\bf x} ~(D_k\psi)^\dagger (D_k\psi) ~~,\cr
P^k &= -i \int d{\bf x} ~\psi^\dagger D_k\psi~~ .\cr}  \eqno(5) $$

To quantize the theory we note that the Chern-Simons field equation
$(\N/ 4\pi) \eps f_{\nu\rho}$$ = j^\mu$
implies  that Chern-Simons fields $a_\mu(x)$ are determined by the matter
field except for non-integrable phases of Wilson line integrals along
non-contractible loops on a torus.   Solving the field equations
in the radiation gauge $div \,{\bf  a}=0$, one finds [4]
$$\eqalign{
a^j(x) &= {\theta_j(t)\over L_j} - {\Phi \over 2L_1L_2} ~\ep^{jk} x_k
  + \ha^j (x)  ~~~, \cr
\ha^j(x) &= {2\pi\over \N} \int d\y ~  \ep^{jk}\nabla^x_k G(\x-\y) ~
  \Big( j^0(y) + {\N\Phi\over 2\pi L_1L_2} \Big)~~~,  \cr
a_0(x) &= - {2\pi \over \N} \int d\y ~ G(\x-\y) ~(\d_1 j^2 - \d_2 j^1)(y)
 ~~~,\cr}     \eqno(6)$$
where
$j^0=\psi^\dagger \psi$,
$j^k = -(i/ 2m) \big\{ \psi^\dagger D_k \psi -
( D_k\psi )^\dagger\psi \big\}$,
and $G(\r)$ is the periodic Green's function on a torus, satisfying
$\Delta G(\r) = \delta(\r) - (1/L_1L_2)$.

The constant parts of $\theta_j$ are non-integrable phases of the Wilson
line integrals.
Residual gauge transformations that respect the boundary
conditions (3) are given by a gauge function
$\Lambda (x) = -\big(  \sum_j 2\pi n_j x_j/ L_j \big) + \tilde{\Lambda} (x)$
where $n_j$'s are integers and  $\tilde\Lambda(x)$ is periodic.
The degree $\tilde\Lambda(x)$
has been made use of to obtain (6), whereas the  rest
of $\Lambda {(x)}$ constitutes large gauge transformations  inducing
$\theta_j \go \theta_j + 2\pi n_j$.   The invariance under large gauge
transformations must be imposed on physical quantities.

In quantum theory,  physical degrees of freedom are the matter fields
$\psi$ (taken to be fermionic), and the non-integrable
phases $\theta_j$'s.  $\psi$'s satify $\{ \psi(\x),\psi^\dagger(\y) \}
=\delta(\x-\y)$ in the fundamental domain of the torus.
$\theta_1$ and $\theta_2$ are canonical conjugate of each
other([4,5]), $[\theta_1,\theta_2]=2\pi i/\N$, as the Lagrangian contains
$(\N/4\pi) (\theta_2 \dot\theta_1- \theta_1\dot\theta_2)$.
In writing the Hamiltonian (5) in terms of these variables with the aid of
(6),  there arises an ordering ambiguity.  We shall
take the ordering of $\psi$ and $\psi^\dagger$ as it is in (5).

The equation of motion for $\theta_j$ is
$$ {\dot\theta}_j = {1\over i}~[\theta_j,H]
   = -\ep^{jk} \, {2\pi\over \N L_k} \, J^k~,   \eqno(7) $$
where $J^k\equiv \int d{\bf x} ~j^k$.  For $\psi(x)$
$$i{\dot\psi} (x) = \Big\{ -{1\over 2m} D_k^2 + a_0(x) + g(x) \Big\}
\psi(x)~,  \eqno(8) $$
where $g(x) = (1/ 2m)   (2\pi/ \N)^2
\int d{\bf y} ~[\nabla_k G({\bf x} - {\bf y}) ]^2  \psi^\dagger\psi (y)$.
Eq.\ (8) differs from the classical equation by the $g(x)$ term [14].    (See
also ref.\ [3] for an analogous result on a plane.)

It is important to recognize that the expression (6) is not completely
equivalent to the Chern-Simons field equations
$(\N/ 4\pi) \eps f_{\nu\rho} = j^\mu$.
 Insertion of (6) into $a_\mu$'s in the equations yields
two nontrivial relations, one Eq.\ (7) and the other
$$ Q + {\N\over 2\pi} \Phi  \approx 0 ~~~, \eqno(9) $$
where $Q= \int d\x~j^0$. $\Phi$ is the flux fixed by the boundary conditions
(3)
with the given $\beta_j(x)$'s, and $Q$ is conserved as a consequence of Eq.\
(8).  Since the relation (9) does not follow from the Hamiltonian and
commutation relations, it has to be imposed  as a constraint.
We have adopted the notation $\approx$ to signify this.

The quantum field theory defined by  (6) -- (9) with the specified ordering of
operators is precisely equivalent to quantum mechanics of anyons [3,14].
This is the reason for our analysing Chern-Simons gauge theory in the present
form.

We now compute commutators of $P^k$ and $H$.  Note that since
$\int d\x~ \psi^\dagger\ha^k\psi = 0$,
$P^k=-i \int d{\bf x} ~\psi^\dagger D_k\psi = -i \int d{\bf x}
{}~\psi^\dagger\Dbar_k\psi$, where
$\Dbar_k \equiv \d_k - i(\theta_k/ L_k) + i(\ep^{kl} x_l \Phi / 2L_1L_2)$.
It follows that
$$[ P^k,\psi(x) ] = i\Dbar_k \psi(x)~~~,~~~
[ P^k, \,\theta_j\, ] = -\ep^{kj} \, {2\pi i\over \N L_k}Q ~~~.
       \eqno(10)$$
In particular, the change in a gauge-invariant operator generated by $P^k$ is a
total derivative.  For instance,
$[P^k,\psi^\dagger\psi(x)] = i \d_k\{\psi^\dagger\psi(x)\}$.   With the aid of
(10) commutators among the operators $P^k$ and $H$ are found to be
$$\eqalign{
[P^j,P^k] &= i \ep^{jk} {2\pi\over \N L_1L_2}\, Q ~\,
   \Big(Q + {\N\over 2\pi} \Phi\Big) ~~~, \cr
[P^j,\,H\,]  &= i \ep^{jk} {2\pi\over \N L_1L_2}\, J^k \,
   \Big(Q + {\N\over 2\pi} \Phi\Big) ~~~, \cr}  \eqno(11)$$
Note that $J^k=P^k/m$ in nonrelativisitc theory.

$P^k$'s and $H$ commute among
themselves only up to the constraint (9).  Hence in the physical Hilbert space
these operators commute, and translation invariance is maintained.
Our conclusion differs from Iengo et al.'s claim [13] that $H$ and $P^k$
commute
among themselves as operators.  Banerjee's analysis on a plane [16] is in
conformity with ours, although a different gauge is chosen.

There are two other sets of important operators,   Wilson
line operators $W_j=e^{i\theta_j}$  and generators of large gauge
transformation $U_j =\exp \big\{ i\ep^{jk} \N \, \theta_k
 -  2\pi i \int d\x ~(x_j/ L_j)\, \psi^\dagger\psi(x) \big\}$
($U_j$ is well defined.  If the coefficient of the integral were a fraction
of $2\pi i$, there would arise inconsistency in $U_j \psi(x) U_j^{-1}$
combined with  (3).)

$U_j$ and $W_j$ satisfy dual relations  $W_1W_2 = e^{-2\pi i/ \N} \,W_2W_1$,
$U_1U_2 = e^{-2\pi i \N}\, U_2U_1$, and  $[W_k, U_j]= 0$.
Also $[U_j,P^k] =$$ [U_j,H]=0$, as $P^k$ and $H$ are gauge-invariant.
Commutator relations among $W_j$, $P^k$ and $H$ are,
however, non-trivial:
$$\eqalign{
[W_j,P^k ] &= \ep^{jk} {2\pi\over \N L_k}  ~Q~ W_j~~,\cr
[W_j, \,H\, ] &= \ep^{jk} {\pi\over \N L_k}
             (J^k W_j + W_j J^k) ~~,\cr}  \eqno(12)  $$
In the nonrelativistic theory $J^k=P^k/m$ so that
 $W_j$'s map an eigenstate into another
eigenstate corresponding to different momemta and energy.

Most of the above results can be directly carried over to Chern-Simons
gauge theory coupled to a Dirac field.   In place of (2) we have
${\cal L}_{\rm matter} =
{i\over 2} \big\{ \psibar \gamma^\mu  D_\mu \psi
- (\overline{ D_\mu \psi }) \gamma^\mu\psi \big\} - m \psibar \psi$.
The current is given by $j^\mu=\psibar \gamma^\mu \psi$.
Gauge-invariant energy-momentum tensors are given by
$T^{\mu\nu}_I={i\over 2} \big\{ \psibar \gamma^\mu D^\nu\psi -
(\overline{ D^\nu \psi}) \gamma^\mu \psi \big\}$.
It follows  that
$$\eqalign{
&H = \int d{\bf x} ~\psibar (-i\gamma^k D_k+ m) \psi ~~, \cr
&P^k = -i \int d{\bf x} ~\psi^\dagger D_k\psi ~~. \cr} \eqno(13)$$
Note that in the Dirac case $J^k$ and $P^k$ are independent quantities.

Most of the relations obtained for the nonrelativisitc case remain valid
for the Dirac case with the substitution
$j^\mu=\psibar \gamma^\mu \psi$ being made.
The only change to be made is the equation for $\psi$:
$i{\dot \psi}= \gamma^0 (-i\gamma^k D_k + m +
    a_0\gamma^0 )\psi(x)$.
The relation (6) and the constraint (9) remain intact.  Direct computations
confirm  (7), (10), and particularly the fundamental algebraic
relations (11) and (12).  $J^k$ is not conserved even in the physical Hilbert
space, however. Therefore $W_j$ no longer
maps an eigenstate of $H$ into  another.

We stress that the relation (11) and (12) are universal.  They are independent
of details of theories.

We now return to the non-relativistic theory and consider the representation of
$P^k$ and $H$ in the corresponding quantum-mechanical anyon system.  The case
of an integer $\N=N$ has been analysed in ref.\ [14].  There are $N$
degenerate vacua $|0_a\ra$ ($a=0, \cdots, N-1$).
$q$-body Schr\"odinger wavefunctions are given by
$\phi_a (\x_1,\ldots,\x_q;t) =(q!)^{-{1\over 2}} \la 0_a | \,\Omega \,
\psi(x_1)\cdots \psi(x_q) |\Psi_q\ra $
where $\Omega =  \exp \big\{ -i\sum_{j=1}^q
 [ (x_1^{(j)} \theta_1 / L_1)   + (x_2^{(j)} \theta_2 / L_2) ]  \big\}$.
The operator $\Omega$ is necessary to insure invariance under large
gauge transformations.  Wavefunctions must have $N$ components as a consequence
of the vaccum degeneracy.  They realize the braid group algebra on a torus.

The representation of $P^k$,  $\hat P^k \, \phi_a \equiv
(q!)^{-1/2} \langle 0_a |\Omega\psi(x_1)\ldots\psi(x_q) P^k |\Psi_q\rangle$,
is found by
permuting  $P^k$ to the left of the $\psi$'s and $\Omega$.  The result is
 simple:
$$\hat P^k = -i\sum_{j=1}^q \nabla_k^{(j)}~. \eqno(14)$$
The Hamiltonian, $\hat H$, is [14]:
$$\eqalign{
\hat H &= - {1\over 2m}
\sum_{j=1}^q \big[ \nabla_k^{(j)}-i \ta^k_{\rm f}(\r_j) \big]^2 ~~,\cr
\ta^k_{\rm f}(\r_j)& = {2\pi\over \N} ~\ep^{kl} \sum_{p \not= j} \Big\{
{1\over 2L_1L_2} ~( x^{(j)}_l - x^{(p)}_l ) +
 \d^{(j)}_l G(\r_j-\r_p)  \Big\} ~~.  \cr}  \eqno(15)$$
It is obvious that all $\hat P^k$ and $\hat H$ commute with each other.
The same conclusion has been reached by Iengo et al.\  [13] in
a different formulation in which the non-integrable phases appear explicitly in
the expressions of $\hat P^k$ and $\hat H$.  Our expressions (14) and (15)
do not contain the $\theta_j$'s.

For the general case where $\N$ is fractional, $\N=N/M$ ($N,M$ coprime), it is
known that there are $NM$ degenerate vacua [7,8].  At present there
are two approaches of interpreting the nature of these vacua.

Following Polychronakos [8],
one might consider, of the $NM$ possible vacua, only $N$ distinct physical
vacua, each having $M$ gauge copies as generated by  $U_j$.  Hence
one  considers a fixed combination of these gauge copies
to represent a physical vacuum.  This can be seen as a sort of ``gauge fixing".
Adopting the same procedure to our case, we will still have a multicomponent
wavefunction, but now $|\Psi_q \rangle$ and $|0_a \rangle$
($a=0,\ldots,N-1$) are linear combinations of $M$ different
gauge-equivalent physical states and vacua, respectively.

The situation is different, however, if we do not regard $U_j$ as operators
of gauge transformation, but instead as physical operators generated by some
physical tunnelling processes.  Such consideration is particularly appropriate
when the model defined by (1) and (2) represents an effective theory of, for
instance, fractional quantum Hall effect (FQHE) (with external magnetic
fields added).   This leads to a conclusion that physical states must
be degenerate as emphasized by Wen and Niu [5].    What we have seen here is
that, not only must physical states be degenerate, but their wavefunctions must
also have multiple components.

To be precise, let us denote the $NM$ degenerate vacua by $|0_{ab}\rangle$
($a=0,\ldots,N-1; b=0,\ldots,M-1$). Noting that $U_1^M$ and $U_2^M$ commute
with
each other, we choose them  to be  eigenstates of $U_i^M$:
$U_i^M |0_{ab}\rangle = e^{i\lambda_i} |0_{ab}\rangle$ ($j=1,2$).  Then in the
$\theta_1$-representation one finds [17] that $u_{ab}(\theta_1)= \la \theta_1 |
0_{ab}\ra  =e^{ib(\lambda_2 + N\theta_1)/M + i \lambda_1 \theta_1/2\pi M } ~
 \delta_{2\pi} [\theta_1 + (\lambda_2- 2\pi Ma)/N]$.
Actions of the $U_i, W_j$ on $|0_{ab}\rangle$ are :
$$\left. \eqalign{
U_1|0_{ab}\rangle &= e^{i(\lambda_1 + 2\pi N b)/M} |0_{ab}\rangle~~,\cr
U_2|0_{ab}\rangle &= e^{i\lambda_2/M} |0_{a,b-1}\rangle~~,\cr }    \right.
\hskip .5cm
\left. \eqalign{
W_1 |0_{ab}\rangle &=e^{-i(\lambda_2 - 2\pi M a)/N} |0_{ab}\rangle~~,\cr
W_2 |0_{ab}\rangle &=e^{i\lambda_1/N} |0_{a-1,b}\rangle~~.\cr}     \right.
            \eqno(16)$$

We require the physical states
$|\Psi\rangle$ to be invariant under $U_i^{-M}$ as well, $U_i^{-M} |\Psi\rangle
= e^{i\omega_i} |\Psi \rangle$.  Since $U_1$ and $U_2$ commute with $U_1^M$ and
$U_2^M$, $U_i^{-1} |\Psi\rangle $ is also a physical state with the same
eigenvalues.   But as $U_1$ and
$U_2$ do not commute, the states generated by them must be degenerate.
It follows from $U_1U_2 = e^{-2\pi iN/M}\, U_2 U_1$ that  $M$ degenerate states
$|\Psi^k\rangle$ ($k=0,\ldots, M-1$) satisfy [5]
$$U_1^{-1} |\Psi^k\ra  = e^{i\nu_k^1} |\Psi^k\ra ~~~,~~~
U_2^{-1} |\Psi^k\ra  = e^{i\nu_k^2} |\Psi^{k+1}\ra ~~~.  \eqno(17) $$

In general, a $q$-particle state $|\Psi^k_q\ra$ can be constructed from the
vacua $|0_{ab}\ra$ satisfying (16) by
$$|\Psi^k_q\ra = {1\over\sqrt{q!}} \sum_{a,b} \int d\x_1\cdots d\x_q~
\phi^k_{ab}
(\x_1,\ldots,\x_q;t)\psi^\dagger(x_q)\cdots\psi^\dagger(x_1)\Omega^\dagger
|0_{ab}\ra /\la 0_{ab} | 0_{ab}\ra~~,     \eqno(18)$$
where $\phi^k_{ab} (\x_1,\ldots,\x_q;t) = (q!)^{-{1\over 2}} \la 0_{ab} |\Omega
\psi(x_1)\ldots\psi(x_q) |\Psi^k_q\ra $.
  To satisfy (17), however,
$|\Psi^k_q\rangle$ can only involve $|0_{ak}\rangle$, {\it i.e.} $b=k$, since
$|0_{ab}\rangle$ pick up diferent phases for different values of $b$ under the
action of $U_1^{-1}$.  Furthermore,   $(i)~\nu^1_k = -(\lambda_1 +
2\pi Nk)/M, ~\nu^2_k=-\lambda_2/M $;  $(ii)~  \phi^k_{ab}=0$ for $b\neq k$ and
$\phi^{k+1}_{a,k+1} = \phi^k_{ak}$ ;  and $(iii)~ \omega_j=-\lambda_j$.  So
states are $M$-fold degenerate, and their wavefunctions take the form of
$(N\times M)$-component matrices $\phi^k_{ab}$ with non-vanishing entries only
in the $k^{\rm th}$ column.

Particularly, in the $\N=1/M$ case, which is of relevance to
FQHE,  elememtary particles have statistics $\theta_s=-M\pi$ and therefore
they are either bosons or fermions depending on whether $M$ is odd or even.
Many-body states
 are nevertheless $M$-fold degenerate, and their wavefunctions have $M$
components.  Implications of these degenerate multicomponent wavefunctions
in the braid group structure of quasi-particles  have yet to
be studied.  In any case, independent of the two approaches just discussed,
$\hat P^k$ and $\hat H$ are represented by (14) and (15), respectively.

\vskip 1. truecm

\centerline{\bf Acknowledgements}

This work is supported in part by the Theoretical Physics Institute
and R.O.C. Grant NSC 81-0208-M032-502 (C.-L.H.) and by the U.S. Department of
Energy under Contract No. DE-AC02-83ER-40105 (Y.H.).  C.-L.H. would like to
thank L. McLerran, M. Voloshin, and the staff of the Theoretical Physics
Insitutute for their hospitality and financial support.

\vfil\eject

\def\ap#1#2#3{{\it Ann.\ Phys.\ (N.Y.)} {\bf {#1}}, #3 (19{#2})}

\def\ijmpA#1#2#3{{\it Int.\ J.\ Mod.\ Phys.} {\bf {A#1}}, #3 (19{#2})}
\def\ijmpB#1#2#3{{\it Int.\ J.\ Mod.\ Phys.} {\bf {B#1}}, #3 (19{#2})}

\def\mplA#1#2#3{{\it Mod.\ Phys.\ Lett.} {\bf A{#1}}, #3 (19{#2})}

\def\plB#1#2#3{{\it Phys.\ Lett.} {\bf {#1}B}, #3 (19{#2})}

\def\np#1#2#3{{\it Nucl.\ Phys.} {\bf B{#1}}, #3 (19{#2})}
\def\prl#1#2#3{{\it Phys.\ Rev.\ Lett.} {\bf #1}, #3 (19{#2})}
\def\prB#1#2#3{{\it Phys.\ Rev.} {\bf B{#1}}, #3 (19{#2})}
\def\prD#1#2#3{{\it Phys.\ Rev.} {\bf D{#1}}, #3 (19{#2})}
\def\prp#1#2#3{{\it Phys.\ Report} {\bf {#1}C}, #3 (19{#2})}
\def\rmp#1#2#3{{\it Rev.\ Mod.\ Phys.} {\bf {#1}}, #3 (19{#2})}

\parindent=15pt

\centerline{\bf References}

\item{[1]} For a review, see eg.:
F. Wilczek, {\it Fractional Statistics and Anyon Superconductivity},
(World Scientific 1990);
E. Fradkin, {\it Field Theories of Condensed Matter Systems},
(Addison-Wesley 1991);
A.L. Fetter, C.B. Hanna, and R.B. Laughlin, \ijmpB {5} {91} {2751};
A. Zee, {\it From Semionics to Topological Fluids},  ITP preprint
NSF-ITP-91-129, (1991);
D. Lykken, J. Sonnenschein,  and N. Weiss, \ijmpA {6} {91} {5155};
S. Forte, \rmp {64} {92} {193};
R. Iengo and K. Lechner,  \prp {213} {92} {179};
D. Boyanovsky,   {\it Gauge Invariance and Broken Symmetries in Anyon
Superfluids}, Pittsburgh preprint PITT-92-01;
Y. Hosotani, {\it Neutral and Charged Anyon Fluids}, Minnesota preprint
UMN-TH-1106/92 (to appear in {\it Int.\ J.\ Mod.\ Phys.} {\bf E}).

\item{[2]}
G. Semenoff, \prl {61} {88} {517};
G. Semenoff and P. Sodano, \np  {328} {89} {753};
T. Matsuyama, \plB {228} {89} {99}; {\it Prog. Theor. Phys.} {\bf 84}
1220 (1990);
S. Forte and T. Joliceur, \np  {350} {91} {589};
D. Boyanovsky, E.T. Newman and C. Rovelli, \prD {45} {92} {1210};
R. Banerjee, \prl {69} {92} {17}.

\item{[3]}
R. Jackiw and S.Y. Pi, \prD {42} {90} {3500}.

\item{[4]}
Y. Hosotani, \prl {62} {89} {2785};  \prl {64} {90} {1691}.

\item{[5]}
X.G. Wen, \prB {40} {89} {7387}; \ijmpB {4} {90} {239};
X.G. Wen and Q. Niu, \prB {41} {90} {9377}.

\item{[6]}
E. Fradkin, \prl {63} {89} {322}.

\item{[7]}
K. Lee, Boston Univ. report, {\it Anyons on spheres and tori},  BU/HEP-89-28;

\item{[8]}
A.P. Polychronakos, \ap{203}{90}{231}; Univ. of Florida preprint,
{\it Abelian Chern-Simons Theories and Conformal Blocks} UFIFT-HEP-89-9;
 \plB {241} {90} {37}.

\item{[9]}
T. Einarsson, \prl {64} {90} {1995};  \ijmpB {5} {91} {675};
 A.P. Balachandran, T. Einarsson, T.R. Govindarajan, and R. Ramachandran,
 \mplA {6} {91} {2801};
T.D. Imbo and J. March-Russel, \plB {252} {90} {84};
Y.S. Wu, Y. Hatsugai and M. Kohmoto, \prl {66} {91} {659};
Y. Hatsugai, M. Kohmoto and Y.S. Wu, \prB {43} {91} {2661}; 10761.

\item{[10]}
S. Randjbar-Daemi, A. Salam, and J. Strathdee, \plB {240} {90} {121}.

\item{[11]}
R. Iengo and K. Lechner, \np {346} {90} {551}; \np {364} {91} {551};
K. Lechner, \plB {273} {91} {463}.

\item{[12]}
K. Lechner,  Trieste SISSA preprint, ``{\it Anyon Physics on the Torus}",
Thesis,
Apr 1991.

\item{[13]}
R. Iengo, K. Lechner, and D. Li, \plB {269} {91} {109}.

\item{[14]}
C.-L. Ho and Y. Hosotani, ``{\it Anyon Equation on a Torus}",
Minnesota preprint UMN-TH-935/91, {\it Int.\ J.\ Mod.\ Phys.} {\bf A} (in
press).

\item{[15]}
Y.-H. Chen, F. Wilczek, E. Witten and B. Halperin, \ijmpB {3} {89} {1001}.

\item{[16]}
R. Banerjee,  ``{\it Energy-Momentum Tensor and Poincare
Algebra in Chern-Simons Theories}", S.N. Bose National Centre for Basic
Sciences preprint, 1992.

\item{[17]} We choose a slightly different basis from that given in [7].
   See also [12].

\vfil

\bye